\documentclass{appolb}
\usepackage{epsfig}
\usepackage{subfigure,amsmath,amssymb}

\begin{document}
\title{The QCD phase diagram in chiral fluid dynamics.%
\thanks{Presented at Excited QCD 2011}%
}
\author{Marlene Nahrgang
and Marcus Bleicher
\address{Institut f\"ur Theoretische Physik, Goethe-Universit\"at, Max-von-Laue-Str.~1, 
60438 Frankfurt am Main, Germany\\
Frankfurt Institute for Advanced Studies (FIAS), Ruth-Moufang-Str.~1, 60438 Frankfurt am Main, Germany}
}

\maketitle
\begin{abstract}
We give a general overview about the approaches to study the phase diagram of QCD. Thereafter, we examine the evolution of a fireball in a chiral fluid dynamic model including nonequilibrium effects.
\end{abstract}
  
\section{Introduction}
At high temperature and densities strongly interacting matter is supposed to form a plasma of quarks and gluons, while at low temperatures and densities it consists of hadronic degrees of freedom. Between these two regimes there must be a phase transition with two aspects, chiral symmetry and confinement.  
For the study of the phase diagram of QCD you have three possible ways to go. 

\begin{figure}
\begin{minipage}{0.49\textwidth}
 \begin{center}
\subfigure{\includegraphics[width=0.8\textwidth]{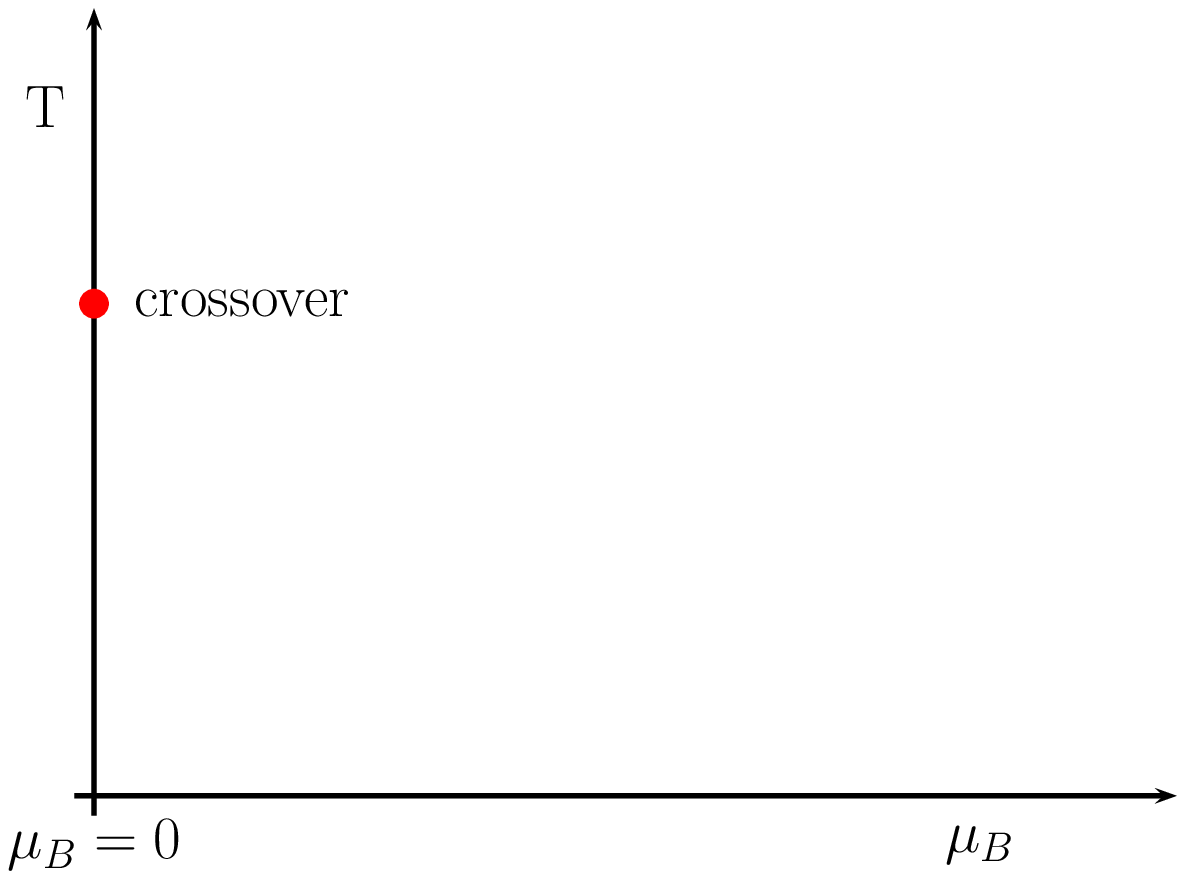}}
\end{center}
\end{minipage}\hfill
\begin{minipage}{0.49\textwidth}
 \begin{center}
 \subfigure{\includegraphics[width=0.8\textwidth]{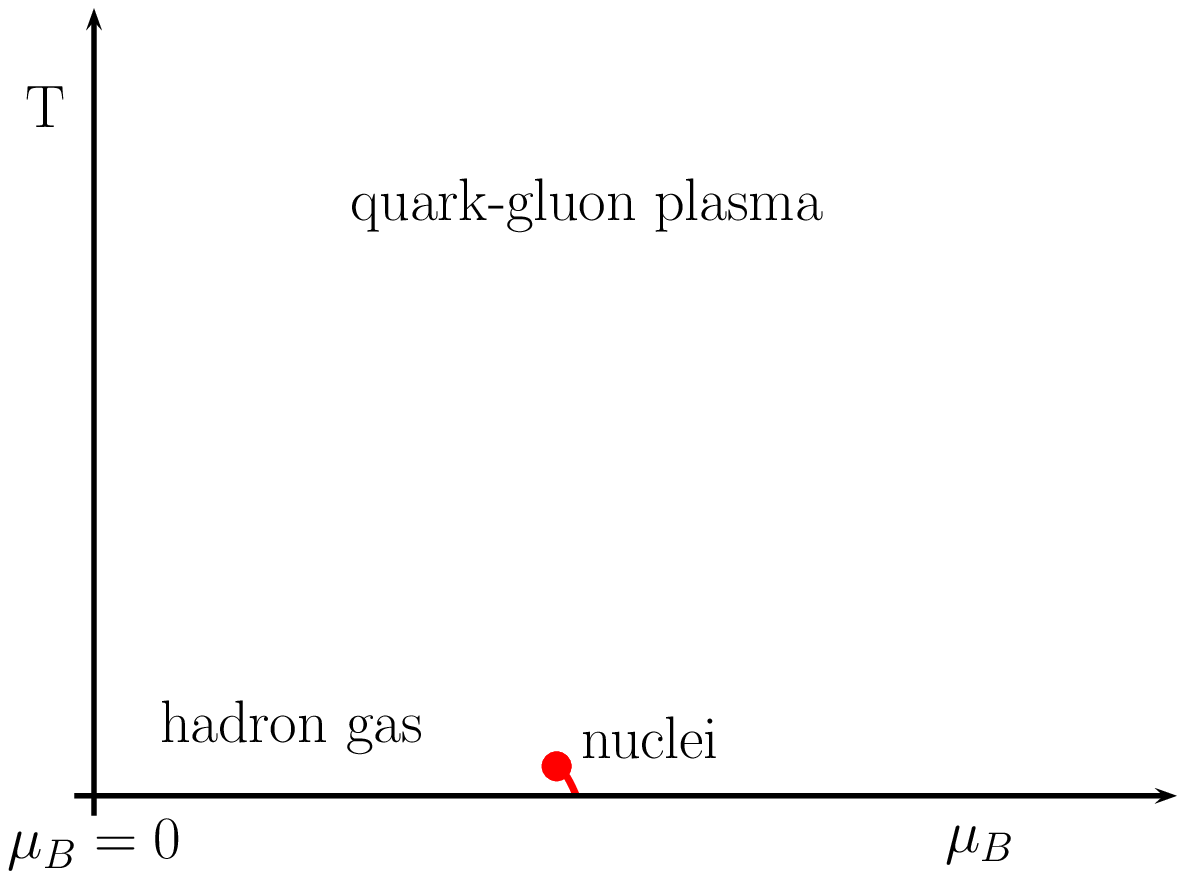}}
\end{center}
\end{minipage}
\caption[Phase diagram from lattice QCD and heavy-ion collisions.]{What we know about the phase diagram of QCD from lattice calculations (left) and from heavy-ion collisions (right).}
 \label{fig:intro_qcdpdlattice}
\end{figure}

\textbf{First}, you are brave and solve the partition function of QCD. This necessarily involves nonperturbative methods, the most promising of which is lattice QCD. With large numerical power QCD is solved on a discretized space-time lattice. This method is, however, only feasible at zero or very small baryonic densities, where it shows that the phase transition is a crossover \cite{Aoki:2006we}. The left plot of figure \ref{fig:intro_qcdpdlattice} shows what we know about the phase diagram from lattice QCD.

 \textbf{Second}, you are strong and collide heavy ions at ultrarelativistic energies. Neutron stars are too far in space and astrophysical observations are too indirect to draw definite conclusions. The Big Bang is too long back in time. You, thus, have to create systems close to the phase transition of QCD in your laboratory. A broad variety of observables has been proposed to study the properties of the quark-gluon plasma and of the QCD phase transition in heavy-ion collisions. Yet, none of them has unambiguously explained the available data. The proposed fluctuation signals for a conjectured critical point  \cite{Stephanov:1998dy,Stephanov:1999zu} have not yet been seen in heavy-ion collisions \cite{Rybczynski:2008cv}. The systems created in a heavy-ion collision are very small, inhomogeneous and expand due to fast dynamics. The finite lifetime of the system was found to limit the growth of the correlation length by critical slowing down \cite{Berdnikov:1999ph}. Real nonequilibrium dynamics might further diminish the critical phenomena but stimulate interesting effects at the first order phase transition \cite{Mishustin:1998eq,Chomaz:2003dz}. The right plot of figure \ref{fig:intro_qcdpdlattice} shows the phase diagram as seen from heavy-ion experiments so far.

\begin{figure}[t]
 \centering
 \includegraphics[width=0.5\textwidth]{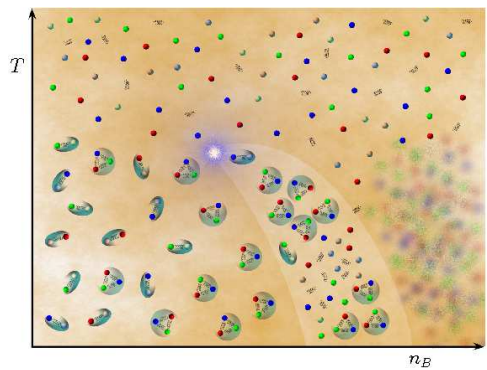}
 \caption[A pictorial view of the phase diagram of QCD inspired by model calculations]{A pictorial view of the phase diagram of QCD inspired by model calculations, which strongly suggest a first order phase transition at higher baryonic densities. In order to see the coexistence region of the first order phase transition the phase diagram is shown in the $T$-$n_B$ plane. The first order phase transition line ends in a critical point.}
 \label{fig:intro_facediagram}
\end{figure}

 \textbf{Third}, you are creative and phenomenologically construct an effective field theoretical model of QCD. Creativity is not unlimited as these models should give a good quantitative description of experimentally measured quantities like cross sections and cover qualitative aspects of the phase diagram, like chiral symmetry and/or confinement \cite{Ratti:2005jh,Schaefer:2007pw}. There are indeed a couple of models that meet these requirements and they can describe certain parameter regions of the phase diagram. These model studies strongly suggest a first order phase transition at high baryonic densities and lower temperatures. This line ends in a critical point. A pictorial view of the phase diagram inspired by model studies is shown in figure \ref{fig:intro_facediagram}.

In the following we present results on a coupled model of the chiral phase transition embedded in a fluid dynamic expansion of quarks \cite{Mishustin:1998yc,Paech:2003fe}. In \cite{Nahrgang:2011mg,Nahrgang:2011mv,Nahrgang:2011ll} chiral fluid dynamic models have been extended by the selfconsistent inclusion of dissipation and noise. 

\section{Chiral fluid dynamics}
Starting from the linear sigma model with constituent quarks \cite{GellMann:1960np} the coupled dynamics of the order parameter of chiral symmetry, the sigma field, and the fluid dynamic expansion of the quarks have been derived \cite{Nahrgang:2011mg}.
The Langevin equation for the sigma mean-field reads
\begin{equation}
 \partial_\mu\partial^\mu\sigma+\frac{\delta V_{\rm eff}}{\delta\sigma}+\eta\partial_t \sigma=\xi\, .
\label{eq:equi_langevineq}
\end{equation}
The effective potential to one-loop level is given by
\begin{equation} 
V_{{\rm eff}}(\sigma, \vec{\pi},T)=U\left(\sigma, \vec{\pi}\right)-2d_q T \int\frac{{\rm d}^3p}{(2\pi)^3}\ln\left(1+\exp\left(-\frac{E}{T}\right)\right) \, ,
\end{equation}
and the damping term is \cite{Greiner:1996dx,Biro:1997va,Nahrgang:2011mg}
\begin{equation}
 \eta=\begin{cases}
        g^2\frac{d_q}{\pi}\left(1-2n_{\rm F}\left(\frac{m_\sigma}{2}\right)\right)\frac{1}{m_\sigma^2}\left(\frac{m_\sigma^2}{4}-m_q^2\right)^{3/2}  & \text{for }m_\sigma>2m_q=2g\sigma_{\rm eq}\\
2.2/{\rm fm}  & \text{for }2m_q>m_\sigma>2m_\pi\\
0 & \text{for }m_\sigma<2m_\pi,2m_q
      \end{cases}
\end{equation}
The stochastic field in the Langevin equation (\ref{eq:equi_langevineq}) has a vanishing expectation value
\begin{equation}
 \langle\xi(t)\rangle_\xi=0\, ,
\end{equation}
and the noise correlation is given by the dissipation-fluctuation theorem
\begin{equation}
 \langle\xi(t)\xi(t')\rangle_\xi=\frac{1}{V}\delta(t-t')m_\sigma\eta\coth\left(\frac{m_\sigma}{2T}\right)\, .
\label{eq:sc_noisecorrelation}
\end{equation}
The pressure of the quarks is locally given by 
\begin{equation}
  p(\sigma, \vec{\pi},T)= -V_{\rm eff}(\sigma, \vec{\pi},T)+U(\sigma, \vec{\pi})\; .
\label{eq:lsm_eos1}
\end{equation}
and the local energy density is obtained from the thermodynamic relation
\begin{equation}
 e(\sigma, \vec{\pi},T)= T\frac{\partial p(\sigma, \vec{\pi},T)}{\partial T}-p(\sigma, \vec{\pi},T)\; .
\label{eq:lsm_eos2}
\end{equation}
In the relativistic fluid dynamic equations we find a source term $S^\nu$ allowing for the energy dissipation from the system to the heat bath
\begin{equation}
\partial_\mu T^{\mu\nu}=S^\nu\, .
\label{eq:fluidT}
\end{equation}

\section{Radial expansion profiles}
\begin{figure}[htb]
 \centering
 \includegraphics[width=0.9\textwidth]{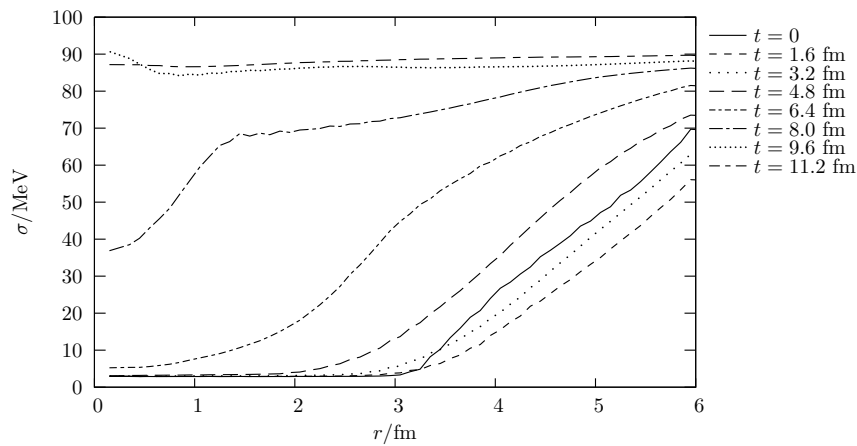}
\caption{The radial profile of the sigma field in a scenario with a first order phase transition.}
 \label{fig:radialproffo.eps}
\end{figure}

\begin{figure}[htb]
 \centering
 \includegraphics[width=0.9\textwidth]{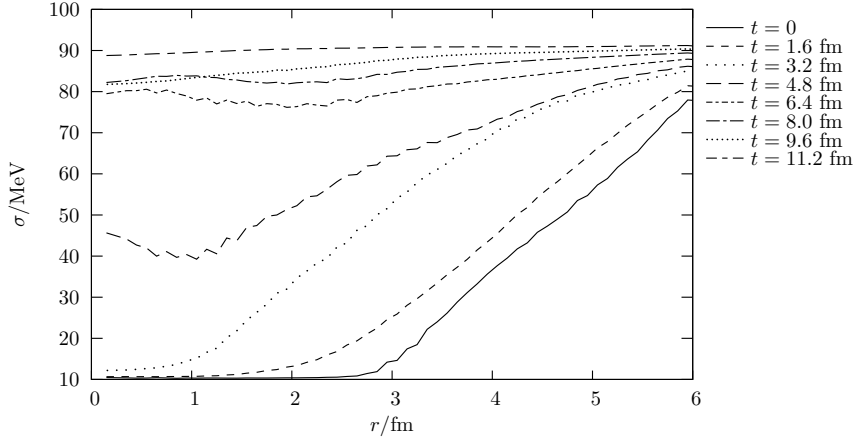}
\caption{The radial profile of the sigma field in a critical point scenario.}
 \label{fig:radialprofcp.eps}
\end{figure}

The system is initialized in an ellipsoidal shape in transverse direction at a maximal temperature of $T=160$~MeV in the inner region. The sigma field is initially in equilibrium with the quark fluid. The expansion of the quark fluid cools the system through the phase transition. In figures \ref{fig:radialproffo.eps} and \ref{fig:radialprofcp.eps} the radial profile of the sigma field is shown at different t

\begin{figure}[htb]

 \centering
 \includegraphics[width=0.9\textwidth]{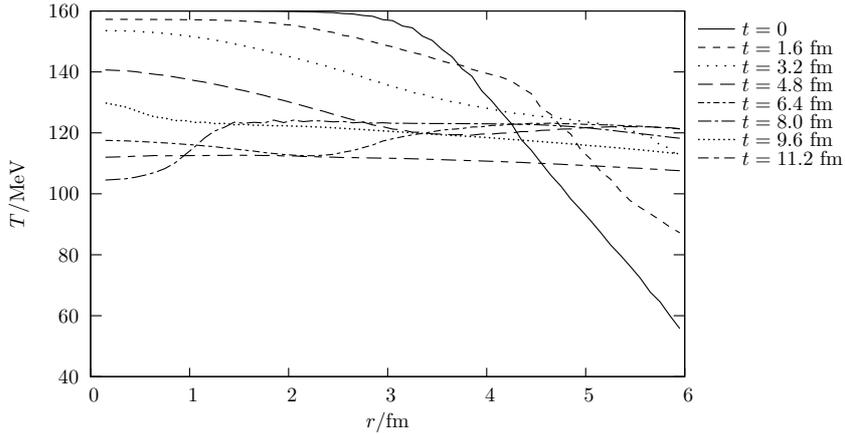}
\caption{The radial temperature profile in a scenario with a first order phase transition.}
 \label{fig:radialproftempfo.eps}
\end{figure}

\begin{figure}[htb]
 \centering
 \includegraphics[width=0.9\textwidth]{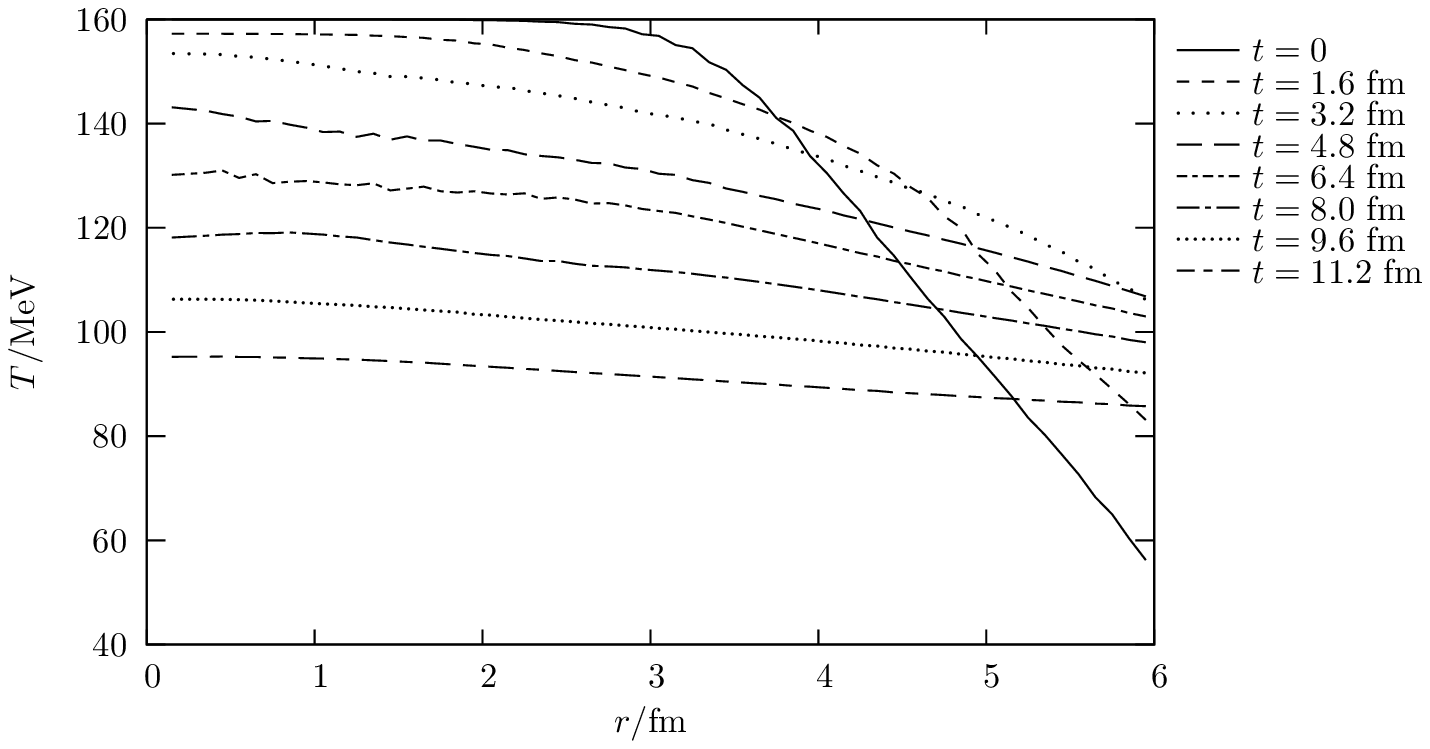}
\caption{The radial temperature profile in a critical point scenario.}
 \label{fig:radialproftempcp.eps}
\end{figure}

Initially the sigma field is close to its high-temperature expectation value $\sigma\simeq0$ in the chirally restored phase. During cooling of the system, the sigma field relaxes. This occurs faster in a critical point scenario, since here $T_c^{\rm cp}=139.88$~MeV than for a first order phase transition $T_c^{\rm fopt}=123.27$~MeV. Moreover, the damping coefficient in a first order phase transition scenario is larger and thus the system is damped strongly into the high-temperature minimum.
Comparing figures \ref{fig:radialproffo.eps} and \ref{fig:radialproftempfo.eps} we see that $t=4.8$~fm the sigma field is supercooled in the inner region. Here, temperatures are already below $T_c^{\rm fopt}$. At larger radii the sigma field relaxes. Due to the dissipation during relaxation the fluid is reheated and thus the temperature is higher at larger radii than in the inner region. For later times the sigma field in the inner region also starts to relax and the temperature rises again. In the inner region the temperature at $t=9.6$~fm is much higher than it was at earlier times between $t\approx6.4\sim8.0$~fm. These two effects, supercooling and reheating, are absent in a critical point scenario.

This work was supported by the Hessian Excellence Inititive LOEWE through the Helmholtz International Center for FAIR. M.N. is supported by the Polytechnische Gesellschaft Frankfurt am Main.

\end{document}